\documentclass[useAMS,usenatbib]{mn2e}
\usepackage{natbib}
\usepackage{graphicx}

\newcommand{\msun}{\mbox{$\,{\rm M}_\odot$}}

\title[The Size Scale of Globular Clusters]{The effect of primordial mass segregation on the size scale of globular clusters}

\author[Haghi et al.]
{Hosein Haghi$^{1}$\thanks{
E-mail:  \mbox{haghi@iasbs.ac.ir} (HH)
\mbox{m.rad@birjand.ac.ir} (SMHR);
\mbox{a.hasani@iasbs.ac.ir} (AHZ);
\mbox{akuepper@astro.columbia.edu} (AHWK)
 }, Seyed Mohammad Hoseini-Rad$^{2}$, Akram Hasani Zonoozi$^{1}$,
\newauthor
Andreas H.W. K\"{u}pper$^{3, 4}$\\\\
$^{1}$Department of Physics, Institute for Advanced Studies in Basic Sciences (IASBS), PO Box 11365-9161, Zanjan, Iran\\
$^{2}$Department of Physics, University of Birjand, PO Box 615/97175, Birjand, South Khorasan, Iran\\
$^{3}$Department of Astronomy, Columbia University, 550 West 120th Street, New York, NY 10027, USA\\
$^{4}$Hubble Fellow
}

\begin{document}

\date{Accepted 2014 August 20. Received 2014 August 20; in original form 2014 June 5}

\pagerange{\pageref{firstpage}--\pageref{lastpage}} \pubyear{2013}

\maketitle

\label{firstpage}

\maketitle

\begin{abstract}
We use direct $N$-body calculations to investigate the impact of primordial mass segregation on the size scale and mass-loss rate of star clusters in a galactic tidal field. We run a set of simulations of clusters with varying degrees of primordial mass segregation at various galactocentric radii and show that, in primordially segregated clusters, the early, impulsive mass-loss  from stellar evolution of the most massive stars in the innermost regions of the cluster leads to a stronger expansion  than for initially non-segregated clusters. Therefore, models in stronger tidal fields dissolve faster due to an enhanced flux of stars over the tidal boundary. Throughout their lifetimes, the segregated clusters are more extended by a factor of about 2, suggesting that (at least) some of the very extended globular clusters in the outer halo of the Milky Way may have been born with primordial mass segregation. We finally derive a relation between star-cluster dissolution time, $T_{diss}$,  and galactocentric radius, $R_G$, and show how it depends on the degree of primordial mass segregation.

\end{abstract}


\section{Introduction}\label{Sec:Intro}

The Milky Way (MW) harbours about 160 globular clusters (GCs; \citealt{Harris96, Harris10}), each containing  up to a few million stars. Observations of GC streams in the MW halo, and simulations of GC systems suggest that the GC populations we observe today are only the very remnants of much richer systems (e.g.~\citealt{Bonaca12, Grillmair13, Brockamp14, Koposov14}). Whether a star cluster survives in the tides of its host galaxy depends crucially on its size: Star clusters with large radii are more susceptible to tidally induced mass-loss, whereas compact systems can survive the depths of galactic centres \citep{Gieles08}.

In the MW, the distances of the GCs from the Galactic Centre range from 0.5 to 125 kpc\footnote{Laevens et al. (2014) and Belokurov et al. (2014) recently announced the discovery of a faint stellar system at 145 kpc distance from the Galactic Centre. However, the nature of this object is presently under debate: while Laevens et al. claim it is a GC, Belokurov et al. suggest it is a faint dwarf galaxy.}.
More than 50\% are found within 10 kpc, but their distribution extends to the very outskirts of our Galaxy. As we expect, most of the clusters in the inner part of the Galaxy are compact, with half-light radii around 3 pc (e.g., \citealt{vandenBerg12}). Initially extended clusters would have been eroded by now and hence would not be observable today. However, the GCs at large Galactic radii on the other hand \textit{all} show large sizes, with half-light radii much larger than 3 pc \citep{Mackey05}. This suggests that GCs were born compact and expanded into their tidal spheres. Understanding this expansion process and what drives it is the motivation for this paper.

Besides the effects from the tidal field, the long-term evolution of GCs is determined mainly by mass-loss due to stellar evolution and stellar dynamics. It is well known that the internal properties  of GCs can undergo significant changes at birth but also during the course of the cluster's dynamical evolution (e.g., \citealt{Heggie03}). It is therefore essential to specify to what extent the present-day properties of GCs, such as their physical sizes and masses are imprinted by early evolution and formation processes and to what extent they are the outcome of long-term dynamical evolution.

$N$-body models show that  in the early evolution of a star cluster there is no balance between the energy flow across the half-mass radius and the production of energy at the centre \citep{Baumgardt02}.
Over its lifetime, it is expected that a GC loses significant amounts of mass or that it may even dissolve entirely. This mass-loss will depend on the physical processes within the cluster (e.g. core collapse, binary formation, stellar evolution), but it will also be sensitive to the  galactic tide within which it orbits (e.g., \citealt{Giersz97, Hurley07,  Heggie08, Gieles11, Brockamp14}). This effect can be amplified by unsteady tides (tidal shocks) when a cluster traverses the Galactic plane, or passes the bulge. Moreover, the most massive GCs may suffer from dynamical friction, which causes them to spiral towards the Galactic Centre.

It is important to investigate how many GCs were disrupted over the lifetime of the Galaxy. As was shown recently by Brockamp et al. (2014), the rate of GC erosion is strongly dependent on the shape and extent of host galaxy potential as well as on initial internal properties of the GCs. The remaining GC around the MW we see today, and their properties may therefore be valuable probes of the Galaxy potential.

The most striking and accessible property of the GCs are their sizes. Recently,  Madrid, Hurley \& Sippel (2012, hereafter MHS12) carried out $N$-body calculations to investigate the physical mechanisms that determine the scale size of star clusters.  They found that the half-mass radius of individual star clusters varies significantly as they evolve
over a Hubble time. Moreover, they showed that it remains constant through several relaxation times, when expansion driven by the internal dynamics of the star cluster and the influence of the host galaxy tidal field balance each other. Therefore, extended GCs  with $r_h >10$ pc are expected to be orbiting at large galactocentric distances. They obtained a relation between the half-mass radius of simulated star clusters orbiting on circular orbits at various galactocentric distances, which takes the mathematical form of a hyperbolic tangent. It should be noted that for all of their models they assumed initially non-segregated distributions of stars.

Mass segregation, the process by which the heavier stars sink towards the centre, and the lighter stars move further away from the centre on a time-scale which is proportional to the relaxation time, is  a natural consequence of two-body relaxation and of the evolution towards energy equipartition in stellar systems. A large number of young clusters with ages significantly smaller than the time needed to produce the observed mass segregation by two-body relaxation alone shows a significant degree of mass segregation, which would probably be primordial and imprinted in a cluster by the star formation processes (for more details, see Sec.~3).

Regardless of the mechanism producing mass segregation, the presence of primordial (or early) mass segregation significantly affects the global dynamical evolution of star clusters. In tidally limited clusters, primordial mass segregation (PMS) leads to a stronger expansion, and hence a larger flow of mass over the tidal boundary. It may therefore help to dissolve them more rapidly. Tidally underfilling clusters, however, can survive this early expansion and have a lifetime similar to that of unsegregated clusters, which has been demonstrated by Vesperini et al. (2009). The authors furthermore  showed that, as the degree of initial mass segregation increases, so does the strength of the initial cluster expansion.
Similarly, Mackey et al. (2007, 2008) demonstrated that the stronger early expansion of mass-segregated clusters, along with the subsequent heating from a population of stellar mass black holes, can explain the radius--age trend observed for massive clusters in the Magellanic Clouds. The degree of primordial, or early, mass segregation is therefore  a crucial parameter in the modelling of GCs.

In this paper we investigate the influence of PMS on the evolution of star clusters in a MW-like potential, in a similar way as MHS12, and compare them with the results of primordially non-segregated models. In Sec.~\ref{Sec:Rh-Rg} we review previous work concerning the size scale--galactocentric distance correlation. In Sec.~\ref{Sec:Evidence} we present some evidence of primordial segregation in GCs. The set-up of our $N$-body models is described in Sec.~\ref{Sec:model}. We compare the results of simulations of non-mass-segregated clusters with primordially mass-segregated clusters in Sec.~\ref{Sec:Results}, and give our conclusions in Sec.~\ref{Sec:Conclusions}.

\section{Half-mass radius--galactocentric distance relation}\label{Sec:Rh-Rg}

The classical notion concerning the size scale of star clusters was that the radius of isolated star cluster remains constant, or changes little over  a few two-body relaxation times \citep{Spitzer72, Lightman78, Aarseth98}. 
But this view was shaken when a clear correlation between the cluster size, $r_h$, and  galactocentric distance, $R_G$, of star clusters was shown in several observational studies (see \citealt{Hodge60, Hodge62} as pioneering studies). Van den Bergh, Morbey \& Pazder (1991) suggested an empirical power-law relation of scale size versus galactocentric distance as $r_h\propto\sqrt{R_G}$ independent of GC classification. However, the data base he was using only included star clusters out to $R_G\simeq30$ kpc.

The correlation between $r_h$ and $R_G$ could be primordial, 
 or it could be the result of  the preferred disruption of large GCs near the Galactic Centre \citep{Vesperini97, Baumgardt03}. Alternatively, it could be due to the expansion of initially small GCs up to the respective Jacobian radius, which is roughly proportional to $R_G^{2/3}$ for a given GC mass.

$N$-body simulations allow us to explore such correlations by detailed dynamical evolution of star clusters in galactic environments, and to determine, which of
these possibilities is more feasible. Based on direct $N$-body simulations of star clusters in a realistic MW-like potential, MHS12 derived a  relationship between scale size and galactocentric distance, and showed that the maximum half-mass radius a star cluster can achieve is proportional to the hyperbolic tangent of the galactocentric distance. The maximum half-mass radius reaches a plateau at large galactocentric distances, which is in contrast with the empirical power-law relation (van den Bergh, Morbey \& Pazder 1991), which does not include this flattening at large galactocentric distances. Moreover, the authors were not able to explain the unusual extent of the outer-halo cluster Palomar\,14, which is located at a Galactocentric distance of 66 kpc and has a half-light radius of 46 pc \citep{Sollima11, Frank14}. Therefore, it is worth investigating the processes driving cluster expansion in more detail.

Changes in the extent of a cluster have to change the energy distribution within the cluster. These energy changes usually go hand in hand with some form of mass-loss. Four different mass-loss processes compete with each other to determine the present-day size of star clusters:
\begin{enumerate}
\item Mass-loss from stellar evolution, which initially increases the cluster size on a short time-scale (e.g., \citealt{Baumgardt03}).
\item Mass-loss driven by two-body relaxation, by which the half-mass radius of a cluster can increase significantly on a relaxation time-scale \citep{Gieles11}.
\item Mass-loss due to tidal stripping, which tends to decrease the size of a cluster \citep{Gnedin99}.
\item Mass-loss from few-body interactions (ejections). However, this has been found to be a subdominant process for clusters in tidal fields \citep{Kuepper08}.
\end{enumerate}
 \citet{Gieles11} demonstrated that, for star clusters evolving at small galactocentric distances, tidal interaction is the dominant process of mass-loss, while for clusters evolving at large galactocentric distances mass-loss is mainly due to internal processes, i.e. stellar evolution and two-body relaxation. However, many of the outer-halo GCs are orbiting in such a weak tidal field that they remain tidally under filling for more than a Hubble time. Their expansion due to two-body relaxation and stellar evolution slows down as stellar evolution becomes less important, the cluster expands, and hence the relaxation time increases. This is why the clusters of MSH12 seem to converge to a certain radius after a Hubble time. In the following, we investigate how the maximum extent a star cluster at a given distance from the Galactic Centre can reach within a Hubble time changes when the clusters are primordially mass segregated.

\section{Evidence of PMS}\label{Sec:Evidence}

In a cluster with a spectrum of stellar masses, the massive stars lose energy via relaxation processes as the cluster moves towards energy equipartition. As a consequence of this process, they segregate towards the central regions, while the lighter stars on average tend to move further away from the centre. This so-called \textit{dynamical mass segregation} is a natural outcome of energy equipartition, and as such happens on a time-scale, which is of the order of several relaxation times. Therefore, it is usually associated with the long-term evolution of clusters. In fact, the majority of Galactic GCs have present-day half-mass relaxation times shorter than their ages. Therefore, they could have established dynamical mass segregation via two-body relaxation.

However, there are some exceptional diffuse outer-halo GCs (e.g., Palomar\,4, Palomar\,14, and AM\,4) with present-day half-mass relaxation times exceeding the Hubble time. Therefore no dynamical mass segregation is, in principle, expected in these clusters. Yet, Frank et al. (2012, 2014) have found clear evidence for mass segregation of main sequence stars in Pal\,4 and Pal\,14. Because of the large two-body relaxation time-scales of these clusters, this could be interpreted as an evidence of PMS. \cite {Zonoozi11,Zonoozi14} have presented a comprehensive set of $N$-body computations of Pal\,14 and Pal\,4 over a Hubble time, and compared the results to the observed mass, half-light radius, flattened stellar mass function and velocity dispersion from \citet{Jordi09} and \citet{Frank12}. They showed that dynamical mass segregation alone cannot explain the mass function flattening in the cluster centre when starting from a canonical Kroupa initial mass function (IMF), and that a very high degree of PMS would be necessary to explain this discrepancy.

There is also observational evidence of PMS in several young Galactic and Magellanic Cloud star clusters with ages shorter than the time needed to produce the observed segregation via two-body relaxation (see e.g., ~\citealt{Hillenbrand97, Bonnell98, Fischer98, de Grijs02, Sirianni02, Gouliermis04, Stolte06, Sabbi08, Allison09, Gouliermis09, de Grijs10}). These star clusters may have formed out of many smaller star-forming clumps.  In each clump, rapid mass segregation may have occurred, sending
the most massive stars to the core of each clump. \citet{McMillan07} showed that when such clumps merge, they will quickly form a virialized cluster, but the mass segregation of the clumps is largely preserved. Alternatively, the PMS in young star clusters could be a result of star formation feedback in dense gas clouds (Murray \& Lin 1996), or due to competitive gas accretion
and mutual mergers between protostars \citep{Bonnell02}.


Another evidence for PMS was proposed by \citet{Baumgardt08} to explain the flattening of the stellar mass functions seen in some GCs, together with the correlation between the slope of the stellar mass function and the cluster concentration that had been discovered by \citet{De Marchi07}.  \citet{Baumgardt08} found that clusters with PMS lose their low-mass stars with a higher rate than non-segregated ones if evolving in a strong  external tidal field, owing to the fact that low mass stars move in the outer parts of the cluster, where they are easily removed by the tidal field. This effect is enhanced if residual gas removal is taken into account, because  the sudden drop of the cluster potential as a result of gas expulsion  leads to preferential loss of low mass stars moving at large radii \citep{Marks08}.

Given the observational evidence for PMS in young star clusters as well as old diffuse GCs, we conclude that at least some, but possibly all, GCs must have started with primordially mass segregation. We therefore aim at shedding light on the effect of PMS on the dynamical evolution, focusing on the size evolution of star clusters by means of direct $N$-body simulations.

\section{$N$-body models}\label{Sec:model}

\begin{figure}
\centering
\includegraphics[width=90mm]{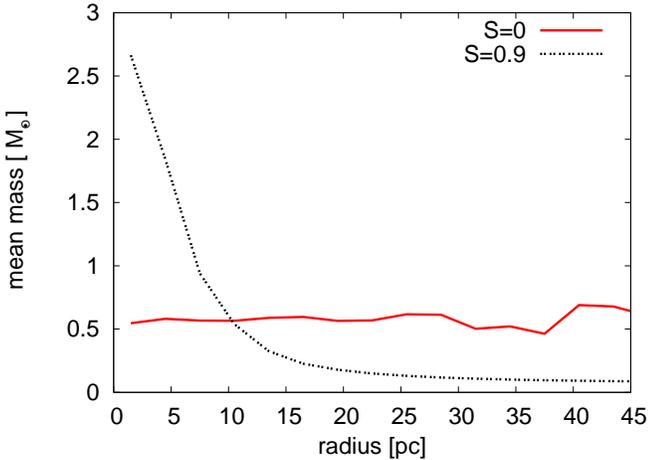}
\caption{ Mean stellar mass as a function of 3D radius for models with (dotted line) and without (solid line) PMS with a
canonical Kroupa IMF, containing  $N=10^5$ stars and an initial half-mass radius of $r_{h,0}=6.2$\,pc. The mass-segregated model is strongly but not entirely segregated, with segregation parameter set to $S=0.9$. The mean mass decreases with increasing distance from the cluster centre for primordially segregated model, while it remains constant through the cluster for model without initial segregation.} \label{meanmass}
\end{figure}

The results presented in this paper are based on a series of
simulations carried out using the state-of-the-art collisional $N$-body code
\textsc{Nbody6} (\citealt{Aarseth03, Nitadori12}). The models were computed on desktop workstations with Nvidia 690 Graphics Processing Units at the Institute for Advanced Studies in Basic Sciences (IASBS).

\textsc{Nbody6} uses a fourth-order Hermite integration scheme with an individual time-step algorithm to progress particles. It invokes regularization schemes to deal with the internal evolution of small-$N$ subsystems, allowing for a detailed handling of binaries and multiples and accounting for close encounters (\citealt{Hurley08a, Hurley08b, Mardling08, Mikkola08, Tout08}).
The code  also includes a comprehensive treatment of stellar evolution by using the \textsc{SSE/BSE} routines and analytical fitting functions developed by Hurley, Pols \& Tout (2000), Hurley, Tout \& Pols (2002), and Hurley et al. (2005).

In our new version of \textsc{Nbody6}, the same as MHS12, we use a three-component analytic galactic tidal field to resemble the MW as described in \citet{Aarseth03} and \citet{Kuepper11}. This galactic potential, $\Phi$, consists of a central
point-mass potential given by
\begin{equation}
\Phi_b (R) = -\frac{G M_b}{R},
\end{equation}
a Miyamoto \& Nagai (1975) disc potential given by
\begin{equation}
\Phi_d (x,y,z) = -\frac{G M_d}{\sqrt{x^2+y^2+\left(a + \sqrt{z^2+b^2}\right)^2}},
\end{equation}
and a logarithmic halo potential of the form

\begin{eqnarray}
\Phi_h (R)= \frac{v_0^2}{2} ~ \ln(R^2+R_c^2).
\end{eqnarray}

Here, $R=x^2+y^2+z^2$ is the distance from the galactic centre at any given time. We use the numerical constants $M_b=1.5 \times 10^{10} M_{\odot}$, $M_d=8.5 \times 10^{10} M_{\odot}$, $a=$4 kpc (disc scalelength) and $b=$0.5 kpc (galactic thickness).  The constant $R_c$ is chosen such that the combined potential of the three components yields a circular velocity of $v_0=220$ km/s in the disk plane at a distance of 8.5 kpc from the galactic centre. All modelled clusters evolve on circular orbits at different galactocentric distances in the disc plane (i.e., the inclination angle of the orbits with respect to the galactic disc is zero).

The initial number of particles for all runs was $N \simeq 10^5$, and the particles were distributed as a Plummer density profile (Plummer 1911),
\begin{equation}
\rho(r) = \frac{3M}{4\pi a^3}\left(1+\frac{r^2}{a^2}\right)^{-5/2},
\end{equation}
where $M$ is the total cluster mass, and $a$ is a scale radius. The half-mass radius,$r_h$, of this profile is related to $a$ by $r_h \simeq 1.305a$. Our models were all set up either with an initial half-mass radius of 6.2 or 3 pc. Hence, the ratio $r_h/r_t$ varies between 0.01 (for cluster with $r_h$=3 pc at $R_G$=100 kpc) and 0.164 (for cluster with $r_h$=6.2 pc at $R_G$=4 kpc). A cluster can be considered as tidally filling if  $r_h/r_t > 0.15$ (Henon 1961).

The models  started with a Kroupa stellar IMF \citep{Kroupa01, Kroupa13}, which consists of two power laws with slope $\alpha=1.3$ for stars with masses, $m$, between 0.08 and $0.5\msun$ and slope $\alpha=2.3$ for more massive stars. The range of stellar masses was chosen to be from 0.08 to 100$\msun$; however we also ran a few models with different upper mass limits, $m_{max}$, to show how it affects the results (see Sec.~\ref{Sec:IC}).  The modelled clusters stars have a metallicity of
$Z = 0.001$ or [Fe/H]$\approx-1.3$.

For segregated systems, mass segregation was set up by using the freely available \textsc{McLuster} code\footnote{\tt https://github.com/ahwkuepper/mcluster} (\citealt{Kuepper11}), which makes use of the segregation routine described in \citet{Baumgardt08}. The degree of segregation can be chosen to be between 0 and 1, where 0 means no segregation and 1 means full segregation, i.e.~ the most massive star sitting in the lowest energy orbit and the second-most massive stars sitting in the second-lowest  energy orbit, and so on. We assigned a quite extreme degree of segregation of $S=0.9$ for all primordially segregated models, in order to show the maximum influence mass segregation can realistically have. The mean mass as a function of radius for this model is shown in Fig.~\ref{meanmass}. The simulations ran for 13 Gyr or until the clusters dissolved.

\section{Results}\label{Sec:Results}

\begin{figure*}
\begin{center}
\includegraphics[width=82mm]{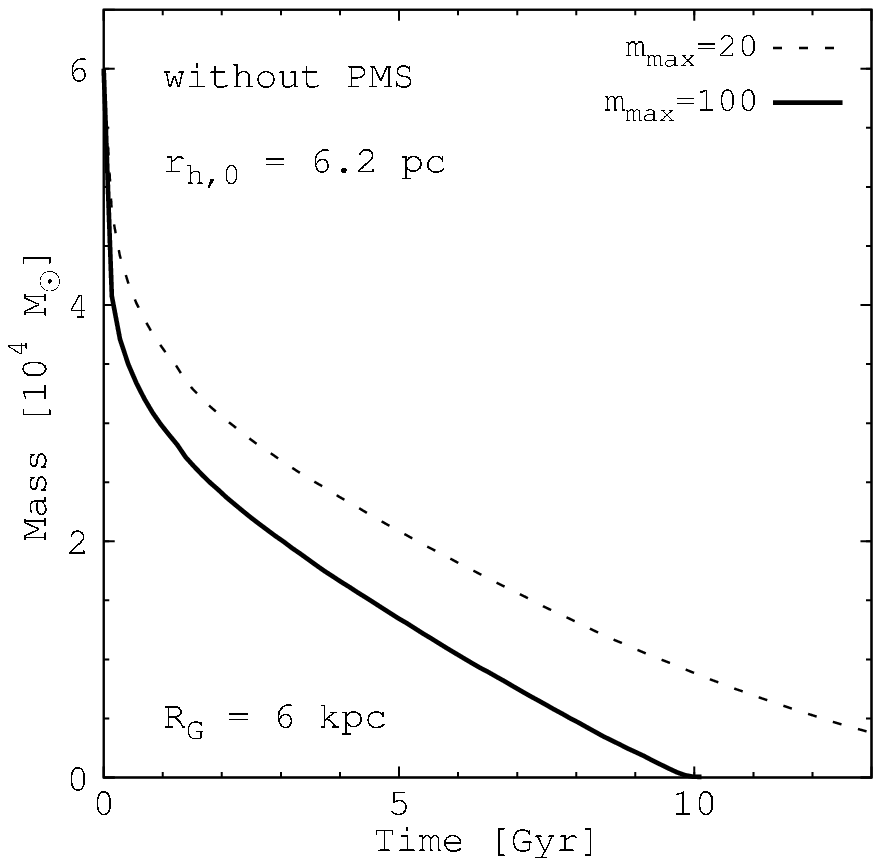}
\includegraphics[width=82mm]{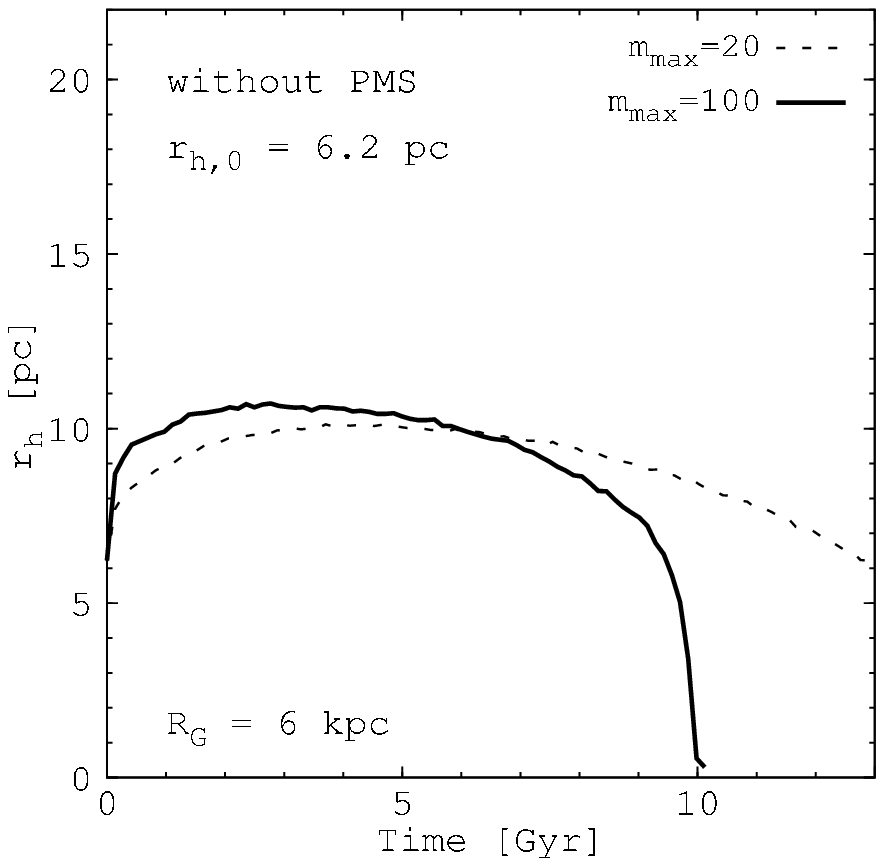}\\
\includegraphics[width=82mm]{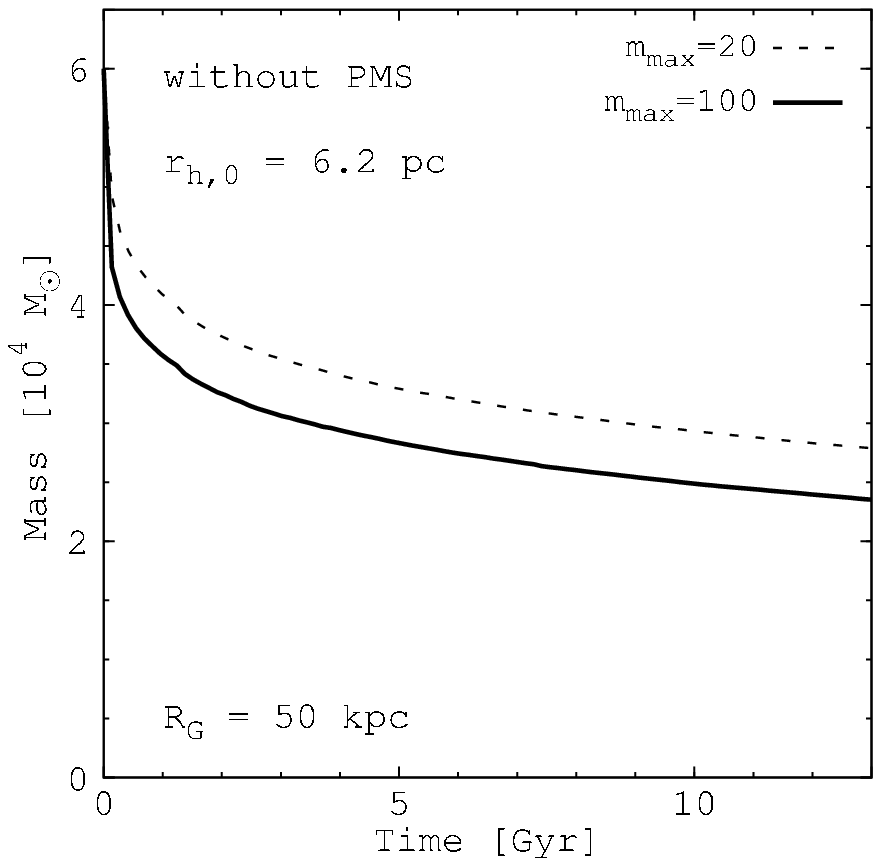}
\includegraphics[width=82mm]{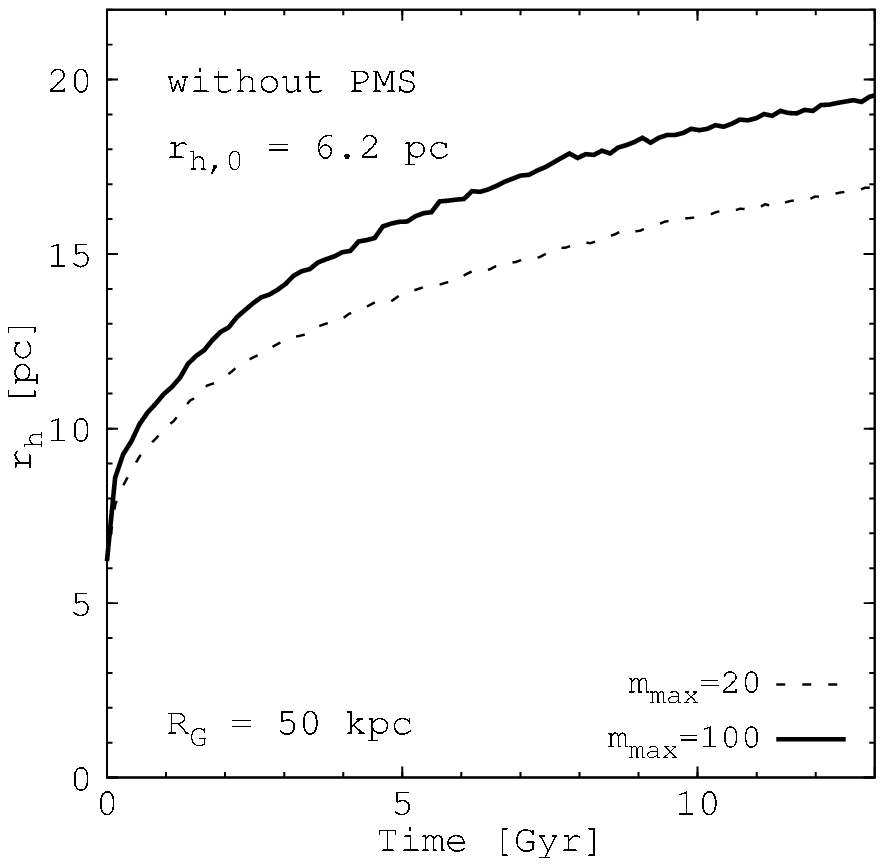}
\caption{The evolution of the total mass (left-hand panels) and half-mass radius (right-hand panels) of the clusters at an orbital
distance of $R_G=6$ (upper panels) and 50 kpc (lower panels) from the galaxy centre for two different upper mass limit
of the stellar IMF.  Models with upper mass limits of 20 and 100 M$_{\odot}$ are indicated by dashed and solid curves, respectively. For the clusters close to the galactic centre, expansion is limited by the strong tidal field, and a higher upper mass means a faster disruption. For the outer-halo clusters, the model with an upper mass limit of 100 M$_{\odot}$ expands to a larger size of 20\% over a Hubble time compared to the 20 M$_\odot$ model.}
\label{uppermass}
\end{center}
\end{figure*}

\begin{figure*}
\centering
\includegraphics[width=85mm]{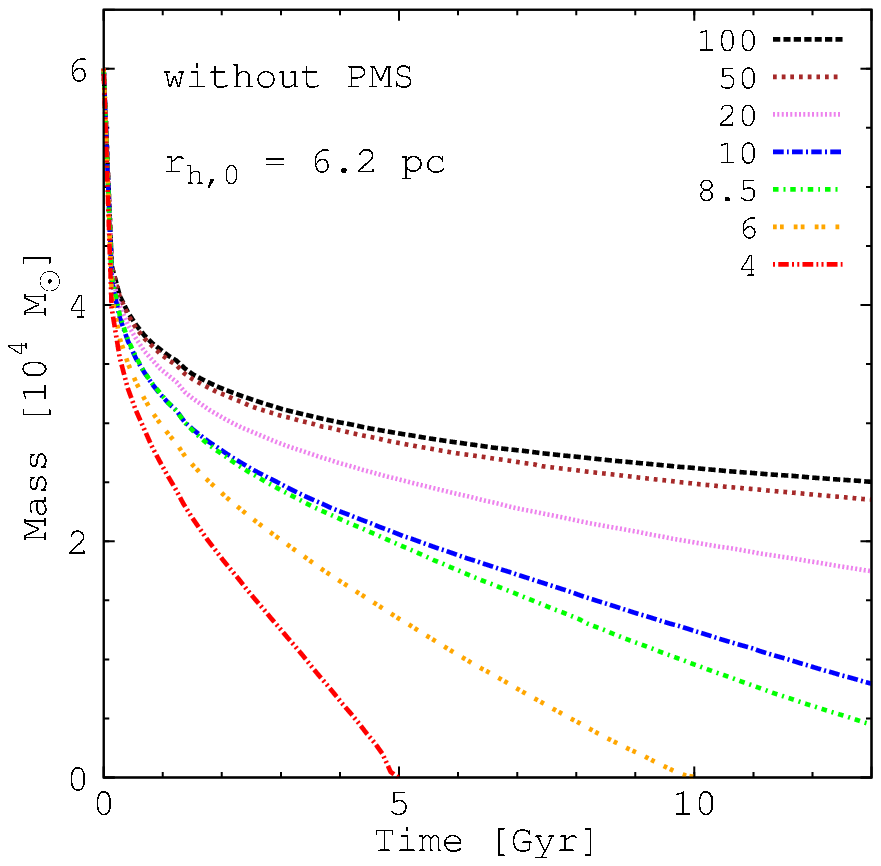}
\includegraphics[width=85mm]{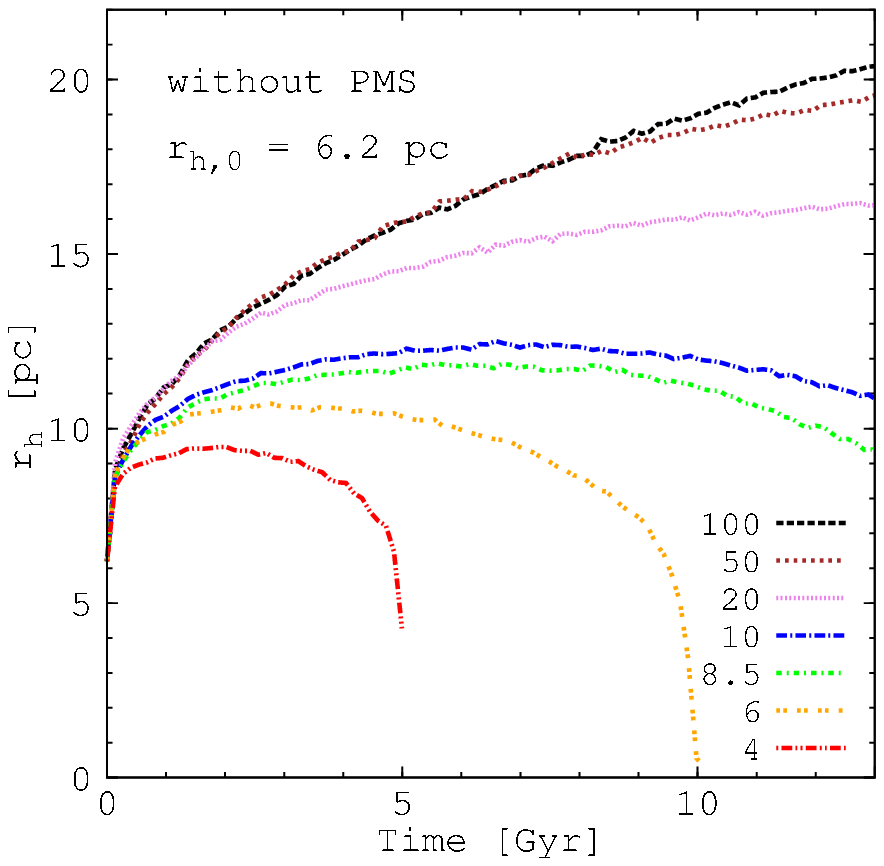}
\caption{ Evolution of initially non-mass-segregated models. All simulations
have an initial 3D half-mass radius of 6.2 pc, and the most massive star has a mass of $m_{max} = 100 M_{\odot}$. Left: the evolution of the total mass for different galactocentric distances, $R_G$. The simulated star clusters evolving at 4 and 6 kpc dissolve before
a Hubble time. Right: evolution of the 3D half-mass radius of the same models. The size of the star cluster located at 10 kpc remains roughly constant for a few Gyr of evolution,  because at this particular galactocentric distance the expansion driven by stellar mass loss and two-body relaxation is balanced by the truncation due to the galactic tidal field.
} \label{mlS0}
\end{figure*}

\subsection{The impact of the stellar mass range}\label{Sec:IC}

Before we start looking into the effect of mass segregation, we examine how changing the initial stellar mass range of a star cluster influences its size scale and dissolution time. This will allow us to make inferences towards the sensitivity of the results of this paper on choosing this crucial initial parameter, which determines the amount of mass that is lost within the first few Myr of a cluster's lifetime. The models in this section are initially non-segregated.

The upper panels of Fig.~\ref{uppermass} show the mass-loss and the evolution of the half-mass radius for two simulations with initial particle numbers of $N=10^5$, both of them with identical initial conditions but with different values of the upper mass limit of the stellar IMF. We show two different models with $m_{max}=20$ and $100\msun$. Both models are evolving  at a galactocentric distance of $R_G=6$ kpc, with the same initial half-mass radius of $r_{h,0}=$ 6.2 pc. Since the heavy stars evolve fast, and hence lose mass almost instantaneously within the first few Myr, the clusters undergo rapid expansion within the first Gyr. This growth is counterbalanced by the flow of stars across the tidal boundary, such that the half-mass radius stops increasing for both models after about 3 Gyr, before it starts decreasing and finally goes to zero as the cluster mass goes to zero. The early expansion is stronger in models with $m_{max}=100\msun$ due to the higher number of heavy stars. But, since the long-term cluster mass-loss for these clusters can be regarded as a runaway overflow over the tidal boundary, the model with $m_{max}=100\msun$ disrupts faster than model with  $m_{max}=20\msun$. Since the clusters are so tidally limited, the model with the higher upper mass limit cannot reach a significantly larger half-mass radius.

In order to evaluate the effect of galactocentric distance, the same simulations were carried out with an orbit at 50 kpc distance from the galactic centre. The lower panels of Fig.~\ref{uppermass} show that, again, the mass-loss from stellar evolution is larger for the cluster with $m_{max}=100\msun$, and hence it expands more than the model with $m_{max}=20\msun$. However, due the large galactocentric distance, the clusters are initially extremely underfilling their tidal sphere. The initial impulsive mass-loss therefore does not cause strong mass-loss in the form of stars flowing over the tidal boundary. Thus, the half-mass radius of the model with $m_{max}=100\msun$ can reach a larger value than the model with $m_{max}=20\msun$ over a Hubble time. The difference between the two half-mass radii is about 20\%. For mass-segregated clusters, this difference will be significantly higher as the extra amount of mass, which is impulsively lost in the first few Myr, is preferentially lost from the centre of the cluster and therefore takes away more binding energy. As we will show in Sec.~\ref{Sec:S}, the model with $m_{max}=100\msun$ can reach a half-mass radius of 33 pc in the same time, when it is initially mass segregated.

\subsection{The impact of PMS}

In order to study the influence of PMS on the size evolution of star clusters in detail, we performed two sets of models, one in which the degree of PMS is zero ($S=0$), and one set initially segregated set with $S=0.9$. Both will be discussed in the following.

\subsubsection{Evolution of the non-segregated models}\label{Sec:S0}

We calculated models orbiting at different galactocentric distances, $R_G=$\,4, 6, 8.5, 10, 20, 50, 100 kpc. All clusters in this section are initially non-segregated, starting with an initial 3D half-mass radius of $\simeq6.2$ pc. We derived the value of 3D half-mass radius of models with NBODY6 taking all stars within the Jacobi radius, and search the radius which contains half of the mass within the Jacobi radius.  The upper mass limit of the stellar IMF in the simulations presented here is $m_{max} = 100\msun$. Fig. \ref{mlS0} shows the mass and half-mass radius evolution of these models. The mass-loss in the  beginning  is the same for all models, because in early stages the evolution is dominated by early impulsive mass-loss associated with stellar evolution of massive stars. All models lose about 40\% of their initial mass during this early evolution. After about first 100 Myr, clusters keep losing mass almost linearly with time. A smaller orbital radius leads to a faster disruption and a smaller half-mass radius after 13 Gyr evolution owing to the enhanced mass-loss driven by the galactic tide, and the stronger cut-off it inflicts on the clusters. The three outermost models appear to have not reached their tidal limit yet and keep expanding till the end of the simulations. The two innermost models (i.e.~$R_G=$ 4, 6 kpc) have lost all of their mass before a Hubble time. Their half-mass radii reach a maximum value, which appears to be clearly linked to their galactocentric distance (as suggested by MHS12), before they decrease again until the cluster dissolves.

\begin{figure}
\centering
\includegraphics[width=85mm]{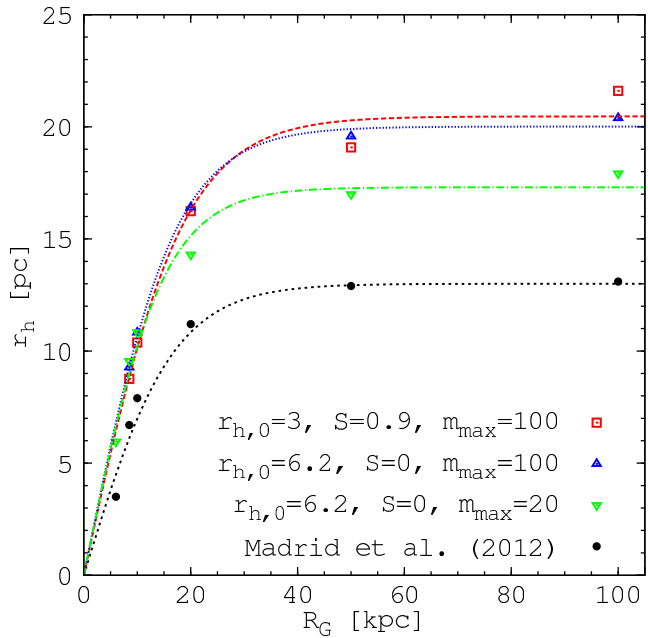}
\caption{Maximum of the three-dimensional half-mass radius of the simulated star
clusters versus galactocentric distances after a Hubble time of evolution. The blue and green triangles depict the values for non-segregated models with the initial 3D half-mass radius of 6.2 pc for two different upper mass limit of the initial stellar mass function. The red squares show the 3D half-mass radius for the primordially segregated runs with initial 3D half-mass radii of 3 pc. The lines are the best-fitting functions given by equation (\ref{eq:rh-RG}). The black dots depict the results of MHS12, for comparison.}
\label{Rh-Rg}
\end{figure}

The maximum value, the half-mass radius can reach throughout a Hubble time at a given galactocentric distance, is shown for all unsegregated models in Fig.~\ref{Rh-Rg} as blue triangles. Also shown are the data points from MHS12 as black dots. We derived a relationship between 3D half-mass radius, $r_h$, of the modelled clusters and galactocentric
distance, $R_G$, in the mathematical form of a hyperbolic tangent, in agreement with the functional form
proposed by MHS12:
\begin{equation}\label{eq:rh-RG}
r_h = r_f\cdot\tanh(a\cdot R_G),
\end{equation}
where $r_f$ and $a$ are two free parameters of the fit. The best-fitting parameters we obtain are $r_f=20.0$\,pc and $a=0.06$\,pc$^{-1}$ for our unsegregated simulations.
It can be seen that the maximum size of our modelled clusters are significantly larger than those of MHS12 by a factor of about 1.5. The inner slope (within the inner 20 kpc where $\tanh(x) \simeq x$) of this function is $a\cdot r_f=1.18$, which is larger than the result of MHS12 by a factor of 1.5.

Where does this difference come from? The initial number of stars ($10^5$), the initial half-mass radii of the clusters (6.2 pc), the metallicity (Z=0.001), and the shape of the IMF (canonical) are the same in both studies. However, a few parameters of the initial set-up of our simulations differ from those assigned by MHS12:
\begin{itemize}
\item First of all, Madrid et al.~use a primordial binary fraction of 5\%. This, however, should rather increase the expansion of the clusters due to the additional heating through the binaries  \citep{Heggie75, McMillan90, Kroupa95}.
\item The simulations of MHS12 also use a slightly different galactic potential. For instance, in our simulations the mass of the disc is $M_{disc}= 8.5\times10^{10}\msun$, while in MHS12 it was $M_{disc}= 5\times10^{10}$\msun, whereas their disc is slightly more concentrated. However, both galactic potentials have a rotational velocity of 220 km/s at $R_G = 8.5$\,kpc, and our star cluster models are also larger at this galactocentric radius.
\item We also assumed that all clusters were moving on a circular orbit in the disc plane, while in MHS12 the initial plane of motion of the star clusters is 22.5 deg inclined to the disc. Madrid, Hurley and Martig (2014) have shown that, for clusters moving near the centre ($<10$ kpc) and in an inclined orbit, the disc shocking at each disc crossing  will strongly enhance the mass-loss rate. However, this effect should become negligible at large galactocentric radii ($\geq10$ kpc) as was shown by Vesperini \& Heggie (1997). So it cannot be the cause for the differences at all galactocentric radii.
\item As discussed in Sec.~\ref{Sec:IC}, the value for the upper mass limit of the stellar IMF is a vital parameter for the expansion of star clusters within a Hubble time. MHS12 use a maximum stellar mass of $m_{max}=50 \msun$, while in the simulations presented here $m_{max}=100\msun$. The difference for our models with  $m_{max}=20$ and $100 \msun$ was up to 20\% for large galactocentric radii (see Fig.~\ref{Rh-Rg}). Hence, this difference will account for part of the differences.
\end{itemize}

Clearly, the relation between maximum half-mass radius and galactocentric distance depends in a systematic way on many of these model parameters. Ideally, the correct choices of model parameters should reproduce observational constraints on this relation. In the inner part, where $a\cdot R_G\ll 1$, it can be seen that $r_h\propto R_G$, which is steeper than expectations from observations of young extragalactic star clusters, which show a considerably weaker dependence, $r_h\propto R_G^{0.1}$.
Moreover, according to equation (\ref{eq:rh-RG}), there is a flattening in the relation between half-mass radius and galactocentric distance: for $R_G\geq$40 kpc, the size scale of the orbiting GCs remains constant at around 20 pc, as the clusters can expand shielded from the truncating effect of the host galaxy's tidal field. Observations of MW GCs, however, give the empirical relation of $r_h\propto R_G^{1/2}$ \citep{vandenBerg91, McLaughlin00}. This is closer to the Roche lobe filling relation, which is less steep with $r_h\propto R_G^{2/3}$ \citep{Scheepmaker07}.

An important model parameter is the initial half-mass radius, as it determines how strong the effect of the impulsive stellar mass loss is. The initial radii of the modelled clusters chosen in this section ($r_{h,0}=6.2$ pc) are motivated through recent finding of \citet{Shin13} who, based on a Fokker-Plank approach, have obtained the best-fit initial size distribution of galactic GCs to be centred around $r_{h,0}=7$ pc. However, such a large value for the initial half-mass radius is in conflict with a recent argument by \citet{Baumgardt10} that most GCs were born compact with sub-parsec size. Also it is in conflict with detailed dynamical modelling of individual clusters, showing that at least some of the clusters must have been born with a several times smaller size. For example, Monte Carlo calculations inferred 0.58, 0.40, and 1.9 pc as initial 3D half-mass radii for M4, NGC 6397, and 47 Tuc, respectively \citep{Heggie08, Giersz09, Giersz11}.

In order to understand whether the PMS helps to reconcile this discrepancy, we calculated a number of models starting with PMS.

\subsubsection{Evolution of the models with primordial segregation}\label{Sec:S}


\begin{figure*}
\centering
\includegraphics[width=80mm]{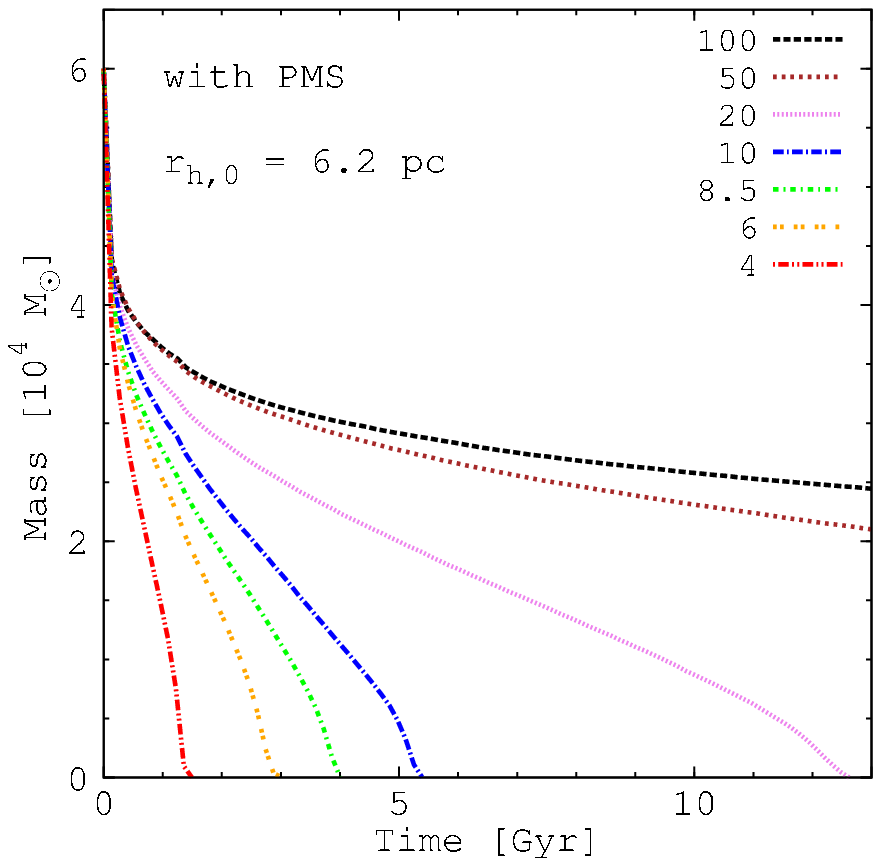}
\includegraphics[width=80mm]{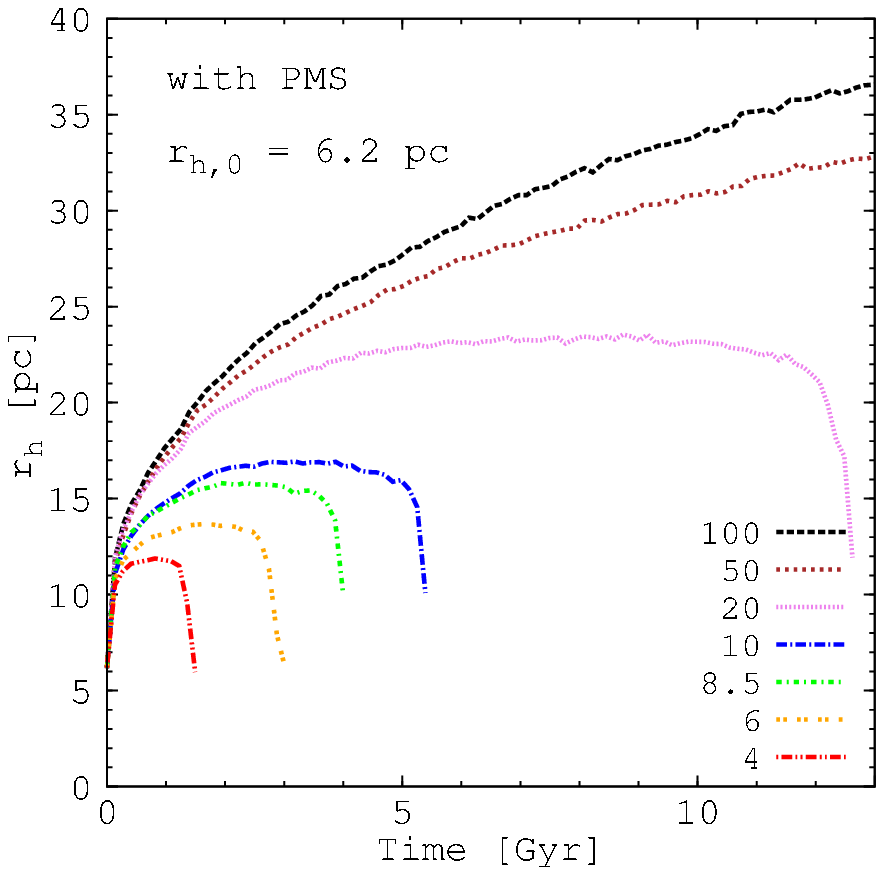}
\caption{ Mass-loss and half-mass radius evolution with time as in Fig.~\ref{mlS0}, but here all models are initially mass segregated. The degree of mass segregation was chosen as $S=0.9$.  In all simulations the upper mass limit of the stellar IMF is $m_{max} = 100 \msun$. All models orbiting at galactocentric distances smaller than about 20 kpc have lost all of their mass before a Hubble time owing to the dominant mechanism of mass-loss by tidal stripping. The half-mass radius of a star cluster located at 20 kpc from the centre of Galaxy remains roughly constant over many Gyr. } \label{mlS9R62}
\end{figure*}

\begin{figure*}
\centering
\includegraphics[width=80mm]{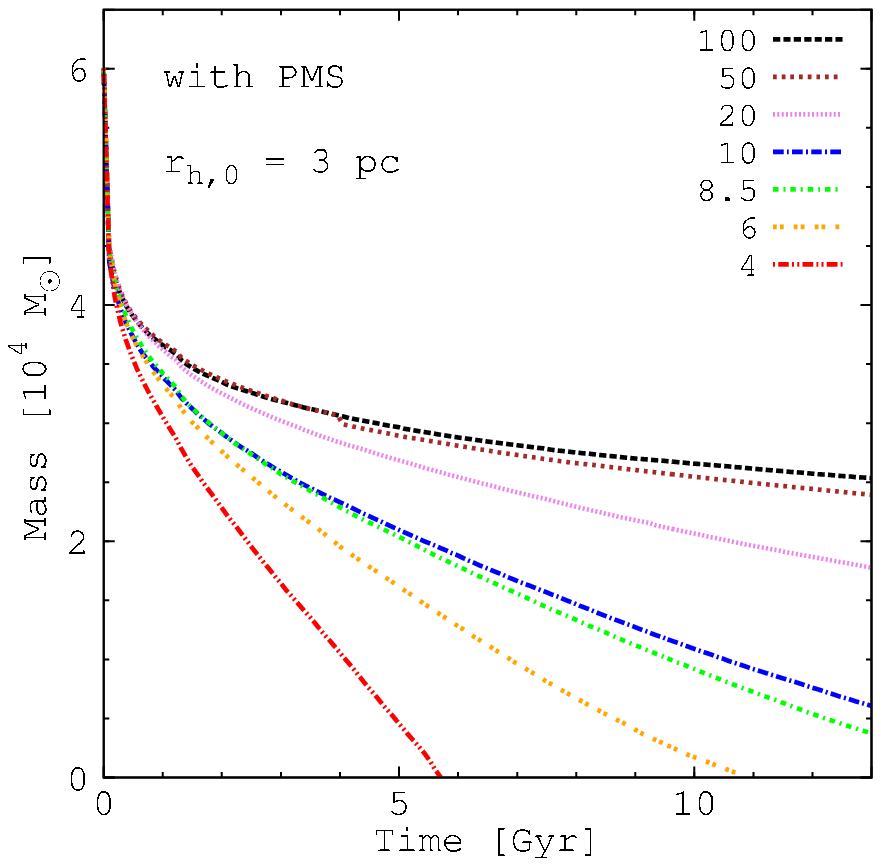}
\includegraphics[width=80mm]{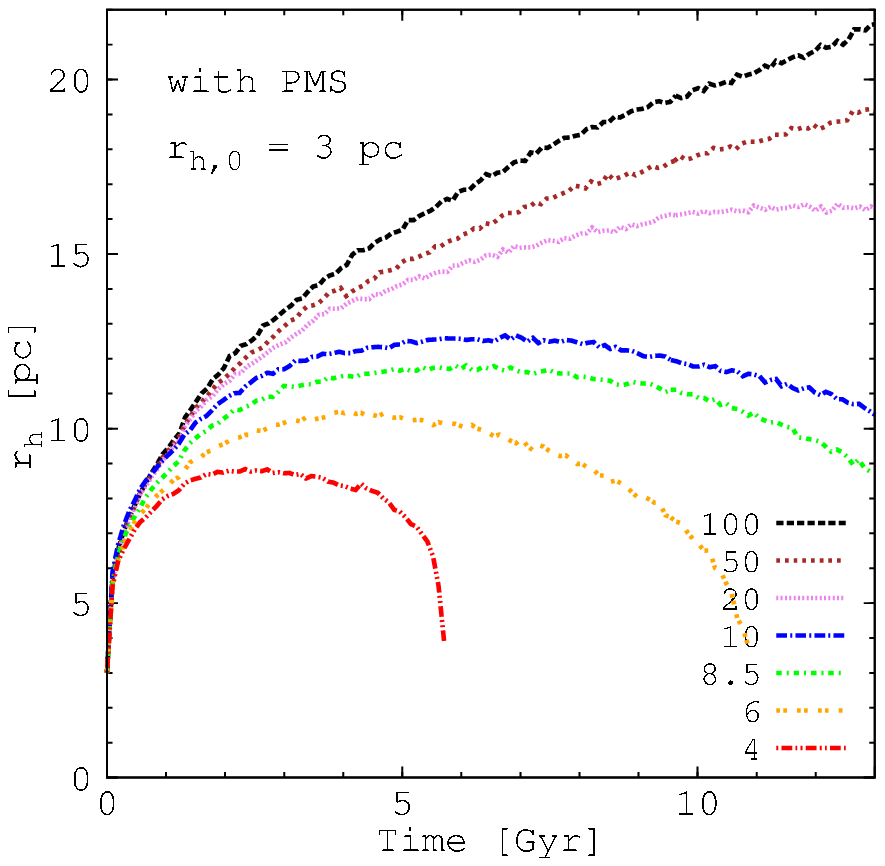}
\caption{ The same as Fig.~\ref{mlS9R62}, but here the initial half-mass radii of the clusters were set to $r_{h,0}=3$\,pc. The evolution of mass and half-mass radii are very similar to the non-segregated models starting with $r_{h,0}=6.2$\,pc, that is, PMS can double the initial cluster expansion from stellar mass loss. } \label{mlS9R3}
\end{figure*}

In this section we describe the results from the simulations of star clusters with PMS and compare them with the  simulations of non-mass-segregated stellar system. In the simulations presented here the most massive star has a mass of $m_{max} = 100 \msun$

For the primordially segregated clusters, the degree of segregation is set to $S=0.9$ in this study. Fig.~\ref{meanmass} shows the initial radial profile of the stellar mean mass for two systems with and without PMS. The decreasing mean mass with increasing distance from the cluster centre  indicates the  mass segregation. The solid line shows the initial mean mass profile of a non-segregated model which remains constant through the cluster. Besides the mass segregation, the clusters start with identical initial conditions as described in Sec.~\ref{Sec:S0} (i.e. initial half-mass radius of $r_{h,0}= 6.2 $ pc and $N=10^5$ stars).

Fig. \ref{mlS9R62} depicts the evolution of the total mass and half-mass radius  at different galactocentric distances, $R_G$, for the initially segregated models. These models undergo a stronger initial expansion, owing to the early impulsive mass-loss associated with stellar evolution which is now happening preferentially deep inside the cluster core, and reach to a larger half-mass radii compared to models starting without primordial segregation. The mass-loss is initially the same for all
models since in this stage it is dominated by stellar evolution of the heaviest
stars. After about 100 Myr, clusters that are closer to the galaxy centre keep losing mass at
a faster pace than the clusters which are further out.

As can be seen, only the star clusters evolving beyond $R_G>20$ kpc survive for a Hubble time. In fact, the early impulsive mass-loss leads to the dissolution\footnote{We define dissolution as the point in time when only 1\% of the initial number of stars is left in the cluster.} of the clusters within the inner 20 kpc. For the survivors, the half-mass radius can reach over five times the initial value after a Hubble time of evolution. Since our models started with 6.2 pc radius, the final half-mass radii are as large as 36 pc, which is larger than the size of most if not all Galactic GCs.

In order to achieve a smaller final radius after a Hubble time, and in order to increase the survival rate in the inner part of the Galaxy, we calculated a set of models with primordial segregation but with smaller initial half-mass radii of $r_{h,0} = 3$\,pc. Comparing Figs ~\ref{mlS0} and ~\ref{mlS9R3}, it can be seen that the final size of these primordially segregated models are very near to those of non-segregated clusters starting  with an initial radius of $r_{h,0}= 6.2$\,pc.

This similarity is also illustrated in Fig.~\ref{Rh-Rg}, where we show the maximum 3D half-mass radius versus galactocentric distance of the simulated star clusters.  The mathematical form of this relation is very similar to that for the non-segregated models from Sec.~\ref{Sec:S0} starting with larger initial half-mass radii.

It seems that a high degree of PMS can make up a factor of 2 in initial compactness of a star cluster.

\subsubsection{Comparison with Galactic GCs}\label{Sec:S}

\begin{figure}
\centering
\includegraphics[width=83mm]{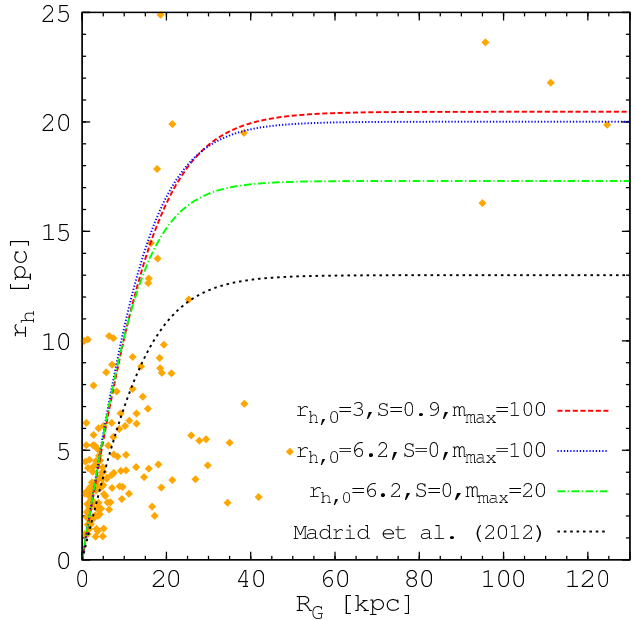}
\caption{3D half-mass radius and galactocentric distance of the MW GC population taken from Harris (2010). Like in Fig.~\ref{mlS9R62}, different lines show the maximum of the 3D half-mass radius of the simulated star
clusters versus galactocentric distances after a Hubble time of evolution.}
\label{Rh-Rg-Harris}
\end{figure}

As an application of our results for the size scale of star clusters, we can compare our results with observations of Galactic globular clusters. We calculated the present-day, three-dimensional half-mass radius of 152 Galactic GCs from the projected half-light radii given in the Harris catalogue (2010).
In order to convert the projected radii, $r_{hp}$ into three-dimensional radii, we used $r_{hp}=\gamma\,r_h$, with $\gamma\approx0.74$ \citep{Spitzer87}. However, if the clusters are mass segregated, which means that the mass-to-light ratio varies with radius, this relation is not accurate and the result has to be taken as rough estimate. In Fig. \ref{Rh-Rg-Harris}, we show how the resulting $r_h-R_G$ relation for different initial condition matches the size of the Galactic GCs. As can be seen, the size of about 10 globular clusters are larger than 13 pc, where our models with different initial conditions can reproduce their size very well.

\subsection{The $T_{diss}-R_{G}$ relation}\label{Sec:relation}

\begin{figure}
\centering
\includegraphics[width=83mm]{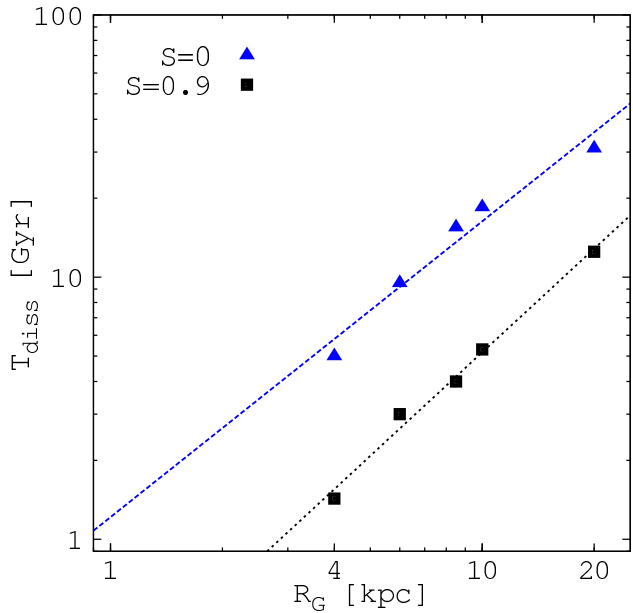}
\caption{ The dissolution time versus galactocentric distance for the primordially segregated (black squares) and non-segregated clusters (blue triangles).
Mass-segregated clusters dissolve very fast in strong tidal fields.  The degree of mass segregation is chosen as $S=0.9$. Power-law fits of the form $T_{diss}\propto R_G ^{\alpha}$ are indicated as dashed lines with the slopes of $\alpha= 1.12\pm0.13$ and $1.31\pm0.08$ for models with $S=0$ and $S=0.9$, respectively. } \label{tdiss}
\end{figure}

Fig. ~\ref{tdiss} depicts the dependence of the cluster dissolution time on the galactocentric distance for primordially segregated and non-segregated clusters.
In agreement with \citet{Vesperini97} and  \citet{Baumgardt03}, the dissolution time of non-segregated clusters increases linearly with galactocentric distance. 
However, even though the amount of impulsive mass-loss from stellar evolution is the same for both segregated and non-segregated clusters, dissolution is faster for the mass-segregated ones. This is due to the larger amount of binding energy which is carried away during stellar evolution from the preferentially centrally located massive stars.

For primordially mass-segregated systems, the initial degree of mass segregation therefore determines the dependence of $T_{diss}$
on $R_G$.  For clusters with $S=0.9$, the amount of binding energy loss due to impulsive mass-loss is larger than those with $S=0$, and the dissolution time is
much shorter than that of the unsegregated systems. We found that the dissolution time scales as
 \begin{equation}
 T_{diss}\propto R_G ^{\alpha(S)},
\end{equation}
where, the exponent $\alpha$, in general, depends on the degree of mass segregation, $S$. For clusters with $S=0.9$, the exponent is $\alpha=1.31\pm 0.08$, while models with $S=0$ have a weaker dependence on $R_G$, with a slope of $\alpha=1.12\pm 0.13$. The slopes marginally agree within the 1$\sigma$ error bars of our fit to the data. However, the offset in $T_{diss}$ is large. It is therefore not possible to derive a unique power law to fit the scaling of $T_{diss}$ with $R_G$ for all values of the $S$ parameter. In other words, for the segregated models the dissolution time is significantly shorter than for the unsegregated systems.

The slope of our $T_{diss}-R_G$ relation for non-segregated models are reasonably close to
that of \citet{Baumgardt03}, who found that $T_{diss}\propto R_G$. However, \citet{Vesperini09} showed that the dissolution times of mass-segregated clusters have a weaker dependence on $R_G$ as $T_{diss}\propto R_G ^{0.43}$. This difference may well be due to the methods how the mass segregated configurations were generated. Maybe more importantly, however, \citet{Vesperini09} investigated tidally filling clusters, whereas our clusters start with a fixed half-mass radius. As discussed above, tidally filling clusters are more susceptible to destruction than tidally under filling clusters, because any cluster expansion causes an increased mass loss. That is, the strong expansion due to mass loss in the initially mass segregated and tidally filling clusters causes them to dissolve quickly, independently of the cluster's galactocentric radius. Hence, the relation found by Vesperini et al.~is close to flat. In our set, the clusters are mostly tidally under filling. Hence they have room to expand as the mass loss from stellar evolution progresses. Thus, the observed relationship between $T_{diss}$ and $R_G$ reflects the weaker influence of the tidal field at larger galactocentric radii.

\section{Conclusions}\label{Sec:Conclusions}

Almost all MW GCs at Galactocentric distances $R_G>40$\,kpc have a larger than average effective radius, i.e. $r_h > 10$ pc.  The presence of these extended clusters is the puzzling issue we addressed here.

Simulations of star clusters without PMS carried out by MHS12 show that, in order to reach such large sizes, the initial sizes of these distant star clusters must have been significantly larger ($r_h>6$ pc) than what is found for most GCs in the MW and in external galaxies, as well as for young massive clusters ($r_h\leq3$ pc).

In this work, using direct $N$-body simulations, we have studied the dynamical evolution of star clusters in the tidal field of the Galaxy starting with and without PMS in order to investigate how a cluster's half-mass radius develops over its lifetime. We showed that the models starting segregated undergo a stronger expansion than the unsegregated ones owing to the rapid, stellar-evolution-induced mass-loss from the inner part of star cluster.

We followed the evolution of clusters at different galactocentric distances, and found that the stronger mass-loss of clusters in the inner 20 kpc of the galaxy limits the expansion of these clusters. For clusters evolving in a weaker tidal field (i.e. at a larger galactocentric distance), PMS leads to significantly larger final size.


Our calculations show that the primordially non-segregated clusters starting with initial three dimensional half-mass radii of 6 pc reach the same size distribution like those clusters that were primordially  segregated but  starting with the smaller radius of 3 pc. Hence, PMS can make up for a factor of 2 in initial cluster size.

We have also explored the dissolution time of evolving star clusters with and without primordial segregation. A larger degree of mass segregation leads to a faster dissolution process due to the stellar-evolution mass-loss from the innermost regions and the stronger expansion. We obtained a relation of the form $T_{diss} \propto R_G^{\alpha}$, where the exponent $\alpha$ depends on the degree of mass segregation. We showed that for initially segregated systems the exponent $\alpha$ is somewhat larger, implying a stronger dissolution for primordially mass-segregated clusters in the outer halo. Such a steepening of the relation with growing galactocentric radius can have significant implications for estimates of the initial number of GCs in galactic GC systems (e.g. Mieske, K\"{u}pper \& Brockamp 2014).

\section*{Acknowledgements}
We would like to thank the referee for constructive comments and suggestions. AHWK acknowledges support through Hubble Fellowship grant HST-HF-51323.01-A awarded by the Space Telescope Science Institute, which is operated by the Association of Universities for Research in Astronomy, Inc., for NASA, under contract NAS 5-26555. This work was made possible by the facilities of Graphics Processing Units at the Institute for Advanced Studies in Basic Sciences (IASBS).

\bsp \label{lastpage}

\end{document}